\documentclass[iop,numberedappendix]{emulateapj-rtx4}
\usepackage{graphicx,natbib,color,bm,url,times,url}
\graphicspath{{./fig/}{./png/}}

\newcommand{\EQ}{\begin{equation}}
\newcommand{\EN}{\end{equation}}
\newcommand{\EQA}{\begin{eqnarray}}
\newcommand{\ENA}{\end{eqnarray}}

\newcommand{\EEq}[1]{Equation~(\ref{#1})}
\newcommand{\Eq}[1]{Equation~(\ref{#1})}

\newcommand{\Sec}[1]{Section~\ref{#1}}

\newcommand{\Fig}[1]{Figure~\ref{#1}}

\newcommand{\Figs}[2]{Figures~\ref{#1} and \ref{#2}}

\newcommand{\Tab}[1]{Table~\ref{#1}}

\newcommand{\bra}[1]{\langle #1\rangle}

{}
{}
{}

{}
{}
{}
{}
{}
{}
{}
{}
{}
{}
{}
{}
{}
{}
{}
{}
{}

{}

{}

{}
{}
{}

\def\Teff{T_{\rm eff}}

\newcommand{\days}{\,{\rm d}}

\newcommand{\dex}{\,{\rm dex}}

\newcommand{\Gyr}{\,{\rm Gyr}}

\newcommand{\yapj}[3]{ #1, {ApJ,} {#2}, #3}

\newcommand{\yan}[3]{ #1, {Astron.\ Nachr.,} {#2}, #3}

\newcommand{\yana}[3]{ #1, {A\&A,} {#2}, #3}

\newcommand{\ygafd}[3]{ #1, {Geophys.\ Astrophys.\ Fluid Dyn.,} {#2}, #3}

\newcommand{\yarep}[3]{ #1, {Astron.\ Rep.,} {#2}, #3}

\newcommand{\ymn}[3]{ #1, {MNRAS,} {#2}, #3}
\newcommand{\ynat}[3]{ #1, {Nature,} {#2}, #3}

\newcommand{\ysph}[3]{ #1, {Solar Phys.,} {#2}, #3}

\newcommand{\sapjE}[2]{ #1, {ApJ}, submitted, #2}

\begin{document}
\title{Enhanced stellar activity for slow antisolar differential rotation?}

\author{Axel Brandenburg$^{1,2,3,4}$ \& Mark S. Giampapa$^5$}
\affil{
$^1$Laboratory for Atmospheric and Space Physics, University of Colorado, Boulder, CO 80303, USA\\
$^2$JILA and Department of Astrophysical and Planetary Sciences, University of Colorado, Boulder, CO 80303, USA\\
$^3$Nordita, KTH Royal Institute of Technology and Stockholm University, Roslagstullsbacken 23, SE-10691 Stockholm, Sweden\\
$^4$Department of Astronomy, AlbaNova University Center, Stockholm University, SE-10691 Stockholm, Sweden\\
$^5$National Solar Observatory, 950 N. Cherry Avenue, Tucson, AZ 85719, USA
}

\submitted{\today, $ $Revision: 1.61 $ $}

\begin{abstract}
High precision photometry of solar-like members of the open cluster M67
with {\em Kepler/K2} data has recently revealed enhanced activity
for stars with a large Rossby number, which is the ratio of rotation
period to the convective turnover time.
Contrary to the well established behavior for shorter rotation periods
and smaller Rossby numbers,
the chromospheric activity of the more slowly rotating stars of
M67 was found to increase with increasing Rossby number.
Such behavior has never been reported before, although it
was theoretically predicted to emerge as a consequence
of antisolar differential rotation (DR) for stars with Rossby numbers
larger than that of the Sun, because in those models the absolute value
of the DR was found to exceed that for solar-like DR.
Using gyrochronological relations and an approximate age of $4\Gyr$
for the members of M67, we compare with computed rotation rates
using just the $B-V$ color.
The resulting rotation--activity relation is found to be compatible with
that obtained by employing the measured rotation rate.
This provides additional support for the unconventional enhancement of
activity at comparatively low rotation rates and the possible presence
of antisolar differential rotation.
\end{abstract}

\email{brandenb@nordita.org}

\section{Introduction}

Main sequence stars with outer convection zones have long displayed
a remarkable universality regarding their dependence of normalized
chromospheric activity on their normalized rotation rate.
This dependence is evident over a broad range of activity indicators
including X-ray, H$\alpha$, and, in particular, the normalized chromospheric
Ca~{\sc ii}~H+K line emission, $R'_{\rm HK}$ \citep[e.g.,][]{Vil84,Noyes84}.
To compare late-type stars of different spectral type, these and other
investigators since then normalized the rotation period $P_{\rm rot}$ by the
star's convective turnover time $\tau$, as determined from conventional
mixing length theory.
This step is obviously model-dependent, but different prescriptions for
$\tau$ as a function of $B-V$ all have in common that $\tau$ increases
monotonically with $B-V$.
With this normalization, the rotation--activity relations of stars of
different spectral type collapse onto a universal curve.
Empirically, the most useful prescription for the function $\tau(B-V)$
is one that minimizes the scatter of $R'_{\rm HK}$ as a function of
$\tau/P_{\rm rot}$, i.e., the {\em inverse} Rossby number.

For $\tau/P_{\rm rot}\ll1$ (slow rotation), the activity indicator
$R'_{\rm HK}$ 
increases approximately linearly with $\tau/P_{\rm rot}$, but
saturates for $\tau/P_{\rm rot}\gg1$.
In this Letter, we focus on a new behavior for values of $\tau/P_{\rm rot}$
that are smaller than what was usually considered in earlier investigations.
In this regime, \cite{GBCSH17} found that $R'_{\rm HK}$ increases
with decreasing values of $\tau/P_{\rm rot}$.
The same trend is reproduced when using the earlier $R'_{\rm HK}$
values of \cite{Giam2006} at somewhat higher spectral resolution where the
effects of color-dependent contamination from the line wings is smaller.
Also calibration uncertainties were shown to be small.

The unconventional scaling of $R'_{\rm HK}$ with $\tau/P_{\rm rot}$
can be associated with a theoretically
predicted increase in {\em differential} rotation (DR) at Rossby
numbers somewhat above the solar value, i.e., for slower rotation
in the normalized sense.
This is the regime of antisolar DR (slow equator, fast poles).
The associated increase of magnetic energy with decreasing rotation
rate was first noticed by \cite{KKKBOP15}; see their Figure~12(b).
The sign reversal of DR, however, has a much longer history and goes
back to early work by \cite{Gil77}.
More recently, with the advent of realistic high-resolution simulations
of solar/stellar dynamos, it became evident that dynamo cycles could
only be obtained at rotation rates that are about three times faster
than that of the Sun \citep{BMBBT11}.
Later, \cite{GYMRW14} found hysteresis behavior in the transition from
solar-like to antisolar-like DR as a function of stellar rotation rate.
Solar-like DR could then be obtained for initial conditions with rapid
rotation.
This led \cite{KKB14} to speculate that the Sun might have inherited
its solar-like DR with equatorward acceleration and slow poles from its
youth when it was rotating more rapidly.
However, subsequent models with dynamo-generated magnetic fields by
\cite{FF14} did not confirm the existence of hysteresis behavior.
Thus, at the solar rotation rate, simulations do indeed produce
antisolar DR.
This is a problem of all solar dynamo simulations to date, but it may be
hoped that the qualitative trends found by \cite{KKKBOP15} would still
hold for the Sun, but at slightly rescaled rotation rates.

The present work supports the prediction by \cite{KKKBOP15} of a
reversed trend in the rotation--activity diagram at very low values
of $\tau/P_{\rm rot}$.
The purpose of this Letter is to compare the new data of \cite{GBCSH17}
with those of other stars, notably those of the Mount Wilson HK project
\citep{Bal+95}\footnote{\url{http://www.nso.edu/node/1335}}.
We focus here particularly on the main sequence stars of \cite{BMM17}
(hereafter BMM) and \cite{SB99} (hereafter SB), for which cyclic
dynamo properties have been analyzed in detail.
Many of those stars have two cycle periods, which fall into one of two
classes in diagrams showing the rotation-to-cycle-period-ratio versus
$R'_{\rm HK}$ or age.
These properties give us a perspective on the stars' evolutionary state
in a broader context.
For the stars of the {\em Kepler} sample of \cite{GBCSH17}, the time
series are still too short, so no information about cyclic activity
exists as yet.
However, based on earlier simulations, we suggest that those stars can
exhibit chaotic variability in $R'_{\rm HK}$ by up to 0.35 dex that
might be detectable over longer time spans.

\begin{table}[t!]\caption{
Sample of solar-like {\em Kepler} stars of \cite{GBCSH17}.
}\vspace{12pt}\centerline{\begin{tabular}{lrccrcccc}
\# & S$\;\;$ & $B$--$V$ & $\Teff$ & $\tau\;$ & $P_{\rm rot}$ & $P_{\rm rot}^\ast$ &
$\!\!\!\!\!\log\bra{R'_{\rm HK}}\!\!\!$ & age \\
\hline
A &  603 & 0.55 & 6091 &  6.4 & $16.6$ & 17.3 & $-4.74$ & 3.7 \\
B &  785 & 0.66 & 5757 & 12.6 & $25.4$ & 24.8 & $-4.82$ & 4.2 \\
C &  801 & 0.68 & 5692 & 13.7 & $20.8$ & 25.7 & $-4.95$ & 2.8 \\
D &  945 & 0.63 & 5856 & 10.8 & $24.3$ & 23.2 & $-4.80$ & 4.3 \\
E &  958 & 0.62 & 5890 & 10.2 & $23.8$ & 22.6 & $-4.89$ & 4.4 \\
F &  965 & 0.72 & 5564 & 15.9 & $26.3$ & 27.4 & $-4.86$ & 3.7 \\
G &  969 & 0.63 & 5856 & 10.8 & $25.7$ & 23.2 & $-5.06$ & 4.8 \\
H &  991 & 0.64 & 5823 & 11.4 & $21.6$ & 23.7 & $-4.84$ & 3.4 \\
I & 1089 & 0.63 & 5856 & 10.8 & $24.5$ & 23.2 & $-4.97$ & 4.4 \\
J & 1095 & 0.61 & 5923 &  9.7 & $22.6$ & 22.0 & $-4.73$ & 4.2 \\
K & 1096 & 0.62 & 5890 & 10.2 & $19.5$ & 22.6 & $-4.86$ & 3.1 \\
L & 1106 & 0.65 & 5790 & 12.0 & $28.4$ & 24.3 & $-4.93$ & 5.3 \\
M & 1212 & 0.73 & 5530 & 16.4 & $24.7$ & 27.8 & $-4.86$ & 3.3 \\
N & 1218 & 0.64 & 5823 & 11.4 & $19.4$ & 23.7 & $-4.78$ & 2.8 \\
O & 1252 & 0.59 & 5988 &  8.5 & $20.3$ & 20.7 & $-4.72$ & 3.9 \\
P & 1255 & 0.63 & 5856 & 10.8 & $24.2$ & 23.2 & $-4.82$ & 4.3 \\
Q & 1289 & 0.72 & 5564 & 15.9 & $23.8$ & 27.4 & $-4.88$ & 3.1 \\
R & 1307 & 0.77 & 5408 & 18.2 & $22.4$ & 29.2 & $-4.95$ & 2.5 \\
S & 1420 & 0.59 & 5988 &  8.5 & $24.8$ & 20.7 & $-4.79$ & 5.5 \\
$\alpha$   &  724 & 0.63 & 5856 & 10.8 & --- & 23.2 & $-4.79$ & $4^\ast$ \\ 
$\beta$    &  746 & 0.67 & 5725 & 13.1 & --- & 25.2 & $-4.89$ & $4^\ast$ \\ 
$\gamma$   &  770 & 0.64 & 5823 & 11.4 & --- & 23.7 & $-4.80$ & $4^\ast$ \\ 
$\delta$   &  777 & 0.63 & 5856 & 10.8 & --- & 23.2 & $-4.90$ & $4^\ast$ \\ 
$\epsilon$ &  802 & 0.68 & 5692 & 13.7 & --- & 25.7 & $-4.95$ & $4^\ast$ \\ 
$\zeta$    &  829 & 0.59 & 5988 &  8.5 & --- & 20.7 & $-4.95$ & $4^\ast$ \\ 
$\eta$     & 1004 & 0.72 & 5564 & 15.9 & --- & 27.4 & $-5.02$ & $4^\ast$ \\ 
$\theta$   & 1033 & 0.57 & 6091 &  7.4 & --- & 19.2 & $-4.74$ & $4^\ast$ \\ 
$\iota$    & 1048 & 0.65 & 5790 & 12.0 & --- & 24.3 & $-5.17$ & $4^\ast$ \\ 
$\kappa$   & 1078 & 0.62 & 5890 & 10.2 & --- & 22.6 & $-4.95$ & $4^\ast$ \\ 
$\lambda$  & 1087 & 0.60 & 5957 &  9.1 & --- & 21.4 & $-4.90$ & $4^\ast$ \\ 
$\mu$      & 1248 & 0.58 & 6025 &  8.0 & --- & 20.0 & $-4.65$ & $4^\ast$ \\ 
$\nu$      & 1258 & 0.63 & 5856 & 10.8 & --- & 23.2 & $-4.90$ & $4^\ast$ \\ 
$\xi$      & 1260 & 0.58 & 6025 &  8.0 & --- & 20.0 & $-4.78$ & $4^\ast$ \\ 
$\pi$      & 1269 & 0.72 & 5564 & 15.9 & --- & 27.4 & $-5.02$ & $4^\ast$ \\ 
$\rho$     & 1318 & 0.58 & 6022 &  8.0 & --- & 20.0 & $-4.73$ & $4^\ast$ \\ 
$\sigma$   & 1449 & 0.62 & 5890 & 10.2 & --- & 22.6 & $-5.13$ & $4^\ast$ \\ 
$\tau$     & 1477 & 0.68 & 5692 & 13.7 & --- & 25.7 & $-4.94$ & $4^\ast$ \\ 
\label{TSum1}\end{tabular}}
\tablenotemark{$\Teff$ is in Kelvin, $\tau$ and $P_{\rm rot}$ is in days,
and age is in $\Gyr$. $P_{\rm rot}^\ast$ (in days) is computed from}
\tablenotemark{\Eq{Pcyc_computed} assuming an age of $t=4\Gyr$,}
\end{table}

\begin{table}[t!]\caption{
F and G dwarfs (italics) and K dwarfs (roman) of BMM.
}\vspace{12pt}\centerline{\begin{tabular}{lrccrrcc}
\# & HD/KIC & $B$--$V$ & $\Teff$ & $\tau\;$ & $P_{\rm rot}$ &
$\!\!\!\log\bra{R'_{\rm HK}}\!\!\!$ & age \\
\hline
{\em a}&     Sun& 0.66 & 5778 & 12.6 & $25.40$ & $-4.90$ & 4.6\\
{\em b}&    1835& 0.66 & 5688 & 12.6 & $ 7.78$ & $-4.43$ & 0.5\\
{\em c}&   17051& 0.57 & 6053 &  7.5 & $ 8.50$ & $-4.60$ & 0.6\\
{\em d}&   20630& 0.66 & 5701 & 12.6 & $ 9.24$ & $-4.42$ & 0.7\\
{\em e}&   30495& 0.63 & 5780 & 10.9 & $11.36$ & $-4.49$ & 1.1\\
{\em f}&   76151& 0.67 & 5675 & 13.2 & $15.00$ & $-4.66$ & 1.6\\
{\em g}&   78366& 0.63 & 5915 & 10.9 & $ 9.67$ & $-4.61$ & 0.8\\
{\em h}&  100180& 0.57 & 5942 &  7.5 & $14.00$ & $-4.92$ & 2.3\\
{\em i}&  103095& 0.75 & 5035 & 17.4 & $31.00$ & $-4.90$ & 4.6\\
{\em j}&  114710& 0.58 & 5970 &  8.0 & $12.35$ & $-4.75$ & 1.7\\
{\em k}&  128620& 0.71 & 5809 & 15.4 & $22.50$ & $-5.00$ & 5.4\\
{\em l}&  146233& 0.65 & 5767 & 12.0 & $22.70$ & $-4.93$ & 4.1\\
{\em m}&  152391& 0.76 & 5420 & 17.8 & $11.43$ & $-4.45$ & 0.8\\
{\em n}&  190406& 0.61 & 5847 &  9.7 & $13.94$ & $-4.80$ & 1.8\\
{\em o}& 8006161& 0.84 & 5488 & 20.6 & $29.79$ & $-5.00$ & 4.6\\
{\em p}&10644253& 0.59 & 6045 &  8.6 & $10.91$ & $-4.69$ & 0.9\\
{\em q}&  186408& 0.64 & 5741 & 11.5 & $23.80$ & $-5.10$ & 7.0\\
{\em r}&  186427& 0.66 & 5701 & 12.6 & $23.20$ & $-5.08$ & 7.0\\
{\rm a}&    3651& 0.84 & 5128 & 20.6 & $44.00$ & $-4.99$ & 7.2\\
{\rm b}&    4628& 0.89 & 5035 & 21.7 & $38.50$ & $-4.85$ & 5.3\\
{\rm c}&   10476& 0.84 & 5188 & 20.6 & $35.20$ & $-4.91$ & 4.9\\
{\rm d}&   16160& 0.98 & 4819 & 22.8 & $48.00$ & $-4.96$ & 6.9\\
{\rm e}&   22049& 0.88 & 5152 & 21.5 & $11.10$ & $-4.46$ & 0.6\\
{\rm f}&   26965& 0.82 & 5284 & 20.1 & $43.00$ & $-4.87$ & 7.2\\
{\rm g}&   32147& 1.06 & 4745 & 23.5 & $48.00$ & $-4.95$ & 6.4\\
{\rm h}&   81809& 0.80 & 5623 & 19.4 & $40.20$ & $-4.92$ & 6.6\\
{\rm i}&  115404& 0.93 & 5081 & 22.3 & $18.47$ & $-4.48$ & 1.4\\
{\rm j}&  128621& 0.88 & 5230 & 21.5 & $36.20$ & $-4.93$ & 4.8\\
{\rm k}&  149661& 0.80 & 5199 & 19.4 & $21.07$ & $-4.58$ & 2.1\\
{\rm l}&  156026& 1.16 & 4600 & 24.2 & $21.00$ & $-4.66$ & 1.3\\
{\rm m}&  160346& 0.96 & 4797 & 22.7 & $36.40$ & $-4.79$ & 4.4\\
{\rm n}& 1653411& 0.78 & 5023 & 18.6 & $19.90$ & $-4.55$ & 2.0\\
{\rm o}&  166620& 0.90 & 5000 & 21.9 & $42.40$ & $-4.96$ & 6.2\\
{\rm p}&  201091& 1.18 & 4400 & 24.4 & $35.37$ & $-4.76$ & 3.3\\
{\rm q}&  201092& 1.37 & 4040 & 25.9 & $37.84$ & $-4.89$ & 3.2\\
{\rm r}& 2198341& 0.80 & 5461 & 19.4 & $42.00$ & $-5.07$ & 7.1\\
{\rm s}& 2198342& 0.91 & 5136 & 22.1 & $43.00$ & $-4.94$ & 6.2\\
1 & 141004 & 0.60 & 5870 &  9.1 & 25.80 & $-5.00$ & 5.6 \\ 
2 & 161239 & 0.65 & 5640 & 12.0 & 29.20 & $-5.16$ & 5.5 \\ 
3 & 187013 & 0.47 & 6455 &  3.1 &  8.00 & $-4.79$ & --- \\ 
4 & 224930 & 0.67 & 5470 & 13.1 & 33.00 & $-4.88$ & 6.4 \\ 
\label{TSum2}\end{tabular}}
\end{table}

\begin{figure*}[t!]\begin{center}
\includegraphics[width=\textwidth]{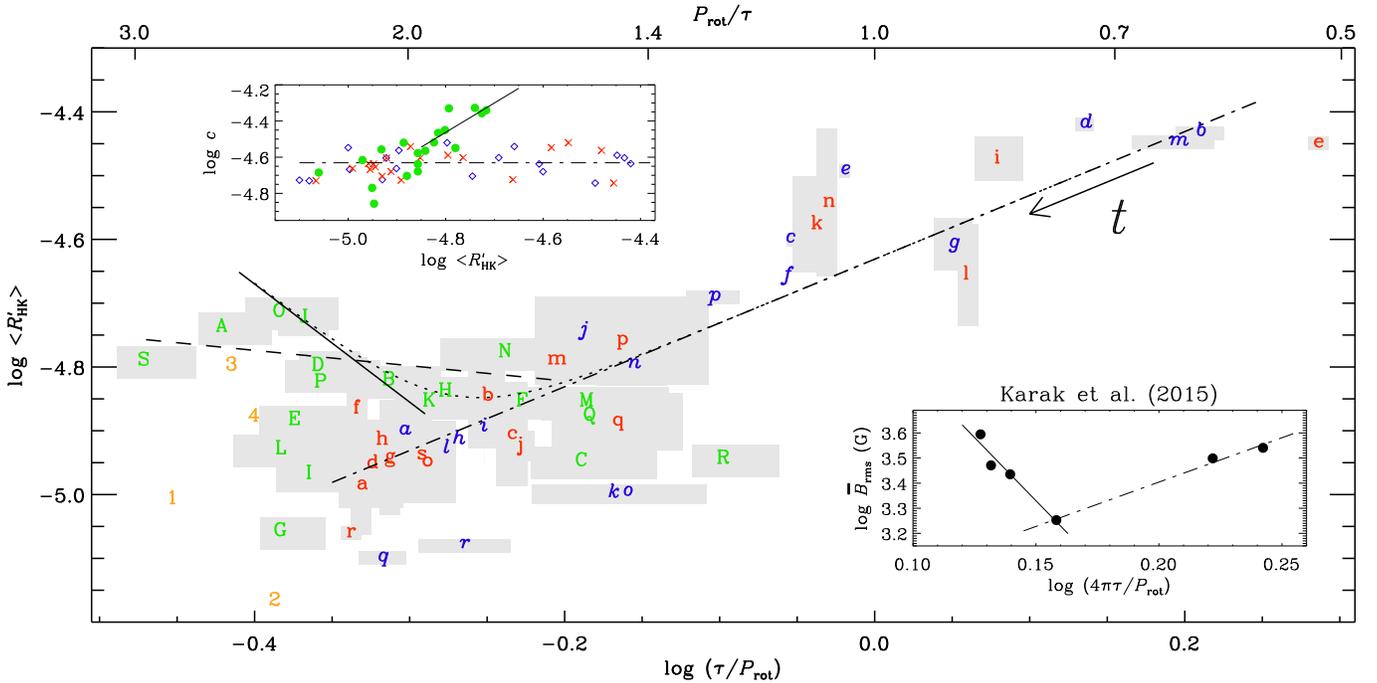}
\end{center}\caption[]{
$\log\bra{R'_{\rm HK}}$ versus $\log (\tau/P_{\rm rot})$ for the
stars of M67 with known rotation periods as green uppercase letters,
the F and G dwarfs of BMM as blue italics characters, the K dwarfs
of BMM as red roman characters, and the four stars of SB with
$P_{\rm rot}/\tau\ge2.4$ as orange numbers 1--4.
On the upper abscissa, the Rossby number $P_{\rm rot}/\tau$ is given.
The dashed-dotted line shows the fit of BMM, whereas the
solid line represents a fit to the residuals in \Eq{cResidual3} for
the nine stars with $\log\bra{R'_{\rm HK}}\ge-4.85$.
The dashed line is a direct fit to the same nine stars and the dotted line
shows the fit given by \Eq{representation}.
The arrow indicates the anticipated evolution with increasing age $t$.
Some of the symbols have been shifted slightly to avoid overlap.
The Sun corresponds to the blue italics $a$.
The upper inset shows the residual $\log c$ versus
$\log\bra{R'_{\rm HK}}$ for the stars of M67 as green filled
circles, the F and G dwarfs of BMM as blue diamonds, and the K dwarfs
of BMM as red crosses.
The lower inset shows the increasing magnetic field strength for small
values of $4\pi\tau/P_{\rm rot}$ from Figure~12(b) of \cite{KKKBOP15}.
}\label{pRHK}\end{figure*}

\section{Representation of the data}

To be able to discuss individual stars in their rotation--activity
diagrams, we denote the stars of M67 by uppercase roman and lowercase
Greek characters and identify them by their Sanders number S in \Tab{TSum1}.
The F and G dwarfs of BMM, represented by lowercase italics characters,
their K dwarfs, indicated by lowercase roman characters, and the
four stars of SB with $P_{\rm rot}/\tau\ge2.4$, indicated by the numbers
1--4, are identified by their HD or KIC numbers in \Tab{TSum2}.
In addition to $B-V$, $P_{\rm rot}$, and $R'_{\rm HK}$, we also give
in both tables the effective temperature $\Teff$ and,
for $B-V>0.495$, the
gyrochronological age $t$ from the relations of \cite{MH08},
\EQ
t=\left\{P_{\rm rot}/ [0.407\,(B-V-0.495)^{0.325}]\right\}^{1.767};
\label{Age}
\EN
see also Equation~(9) of BMM.\footnote{
This relation gives 3\%--14\% smaller ages than the one of \cite{Bar10},
which was also used by \cite{GBCSH17}, taking $\tau$ from \cite{BK10}.
Here we use \Eq{Age} for consistency with BMM.}
\EEq{Age} can be inverted to compute instead $P_{\rm rot}$ under
the reasonable assumption that $t=4\Gyr$ is valid for all stars of M67;
evidence comes from isochrones \citep{Sara2009,Onehag2011}, gyrochronology
\citep{Barnes2016}, and chromospheric activity combined with gyrochronology
\citep{GBCSH17}.
This yields
\EQ
P_{\rm rot}^\ast=0.407\,(B-V-0.495)^{0.325}\;t^{0.565},
\label{Pcyc_computed}
\EN
where the asterisk is used to distinguish the computed value
from the measured one.
Next, using the semi-empirical relationship for $\tau(B-V)$ of
\cite{Noyes84} in the form
\EQ
\log\tau=1.362-0.166x+0.03x^2-5.3x^3,
\label{BV}
\EN
with $x=1-(B-V)$ and for $B-V<1$, we obtain $\tau/P_{\rm rot}^\ast$ as
a monotonically increasing function of $B-V$ in the range from $0.55$
to $0.8$.

Given these relations, we first show in \Fig{pRHK} all stars
with measured rotation periods in the rotation--activity diagram.
Error bars in $\bra{R'_{\rm HK}}$ and $P_{\rm rot}$ are marked by gray boxes.
The stars of BMM follow an approximately linear increase that can be described
by the fit $\log\bra{R'_{\rm HK}}\approx\log(\tau/P_{\rm rot})+\log c$, where
$\log c\approx-4.63$.
However, in spite of significant scatter, there is a clear increase
in activity for most of the stars of the sample of M67 as
$\tau/P_{\rm rot}$ decreases.
HD~187013 and 224930 (orange symbols 3 and 4 with $P_{\rm rot}/\tau=2.6$
and $2.5$, respectively) of the Mount Wilson stars are found to be compatible
with this trend.
We show two separate fits in \Fig{pRHK}, a direct one and one that has
been computed from a fit to the residual between $\log\bra{R'_{\rm HK}}$
and $\log (\tau/P_{\rm rot})$, i.e.,
\EQ
\log\bra{R'_{\rm HK}} - \log (\tau/P_{\rm rot})
=\log c_1 + \rho\log\bra{R'_{\rm HK}}.
\label{cResidual}
\EN
In the upper inset of \Fig{pRHK} we denote this residual by $\log c$,
where $c$ is a function of $\bra{R'_{\rm HK}}$.
\EEq{cResidual} is then written in terms of an expression for
$\log\bra{R'_{\rm HK}}$ versus $\log(\tau/P_{\rm rot})$.
The parameters in \Eq{cResidual} have been computed from the nine out
of 19 stars for which $\log\bra{R'_{\rm HK}}\ge-4.85$.
This yields $\log c_1\approx2.92$ and $\rho\approx1.54$,
which is shown in the upper inset of \Fig{pRHK} as a solid line.\footnote{
\cite{GBCSH17} computed $\log c_1$ and $\rho$ for all 19 stars using
$\tau(B-V)$ from \cite{BK10} instead of \cite{Noyes84}; their values are
therefore somewhat different: $\log c_1\approx1.11$ and $\rho\approx1.25$.}
Solving for $\log\bra{R'_{\rm HK}}$ gives
\EQ
\log\bra{R'_{\rm HK}} = \log c_2 + \mu_2 \log (\tau/P_{\rm rot}),
\label{cResidual3}
\EN
where $\log c_2=\mu_2\log c_1\approx-5.41$ with
$\mu_2=(1-\rho)^{-1}\approx-1.85$.
It is shown in the main part of \Fig{pRHK} as a solid line.
By comparison, the direct fit for the same nine stars gives
$\log c_2^\ast\approx-4.87$ and $\mu_2^\ast=-0.24$ and is shown in
\Fig{pRHK} as a dashed line.
In addition, we combine the fit of BMM with that of \Eq{cResidual3} as
\EQ
\bra{R'_{\rm HK}}= 
\left\{\left[c_0\,(\tau/P_{\rm rot})\right]^q+
\left[c_2\,(\tau/P_{\rm rot})^{\mu_2}\right]^q\right\}^{1/q},
\label{representation}
\EN
where $c_0=10^{-4.631}$ is the residual of BMM and $q=5$
is chosen large enough to make the transition between
the two fits sufficiently sharp.
This special representation now applies to the whole range of
$\tau/P_{\rm rot}$ and we return to it in \Sec{Evolution}.
To remind the reader of Figure~12(b) of \cite{KKKBOP15}, we show
in the lower inset of \Fig{pRHK} the magnetic field strength versus
$4\pi\tau/P_{\rm rot}$.
The $4\pi$ factor emerges because in those models, rotation is
controlled by the Coriolis force, which is proportional to $2\Omega$,
where $\Omega=2\pi/P_{\rm rot}$ is the angular velocity.

\begin{figure*}[t!]\begin{center}
\includegraphics[width=\textwidth]{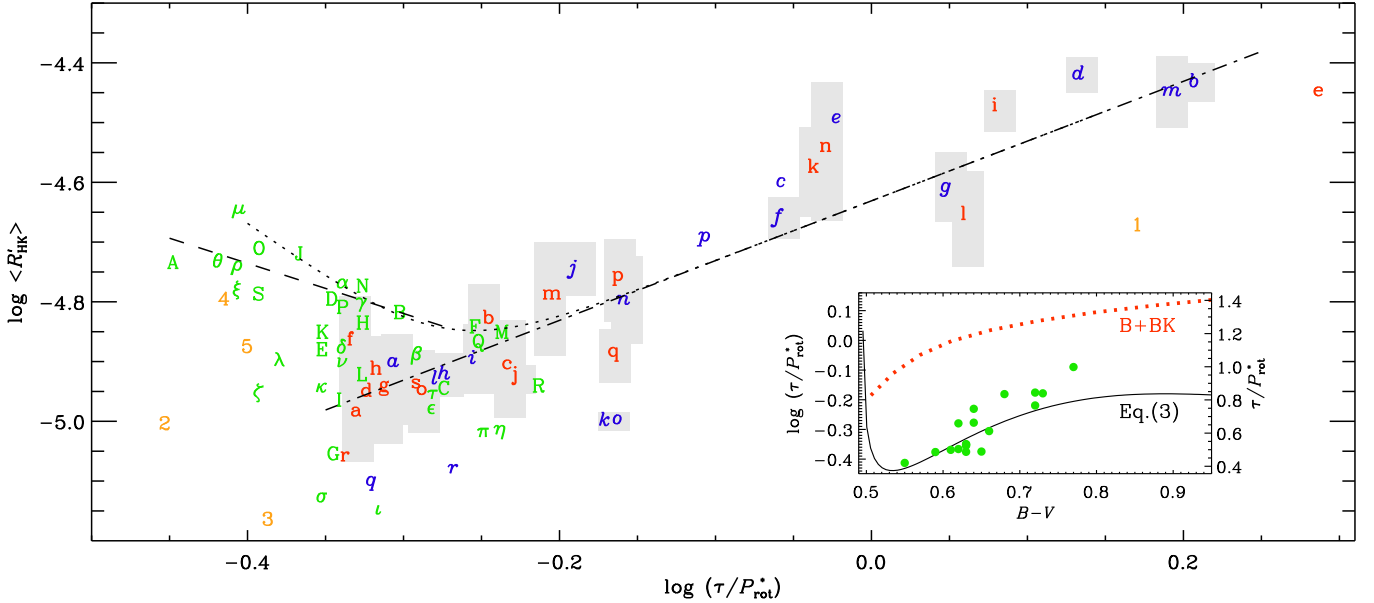}
\end{center}\caption[]{
Similar to \Fig{pRHK}, but now with rotation periods computed from $B-V$
using \Eq{Pcyc_computed} and the assumption that M67 is $4\Gyr$ old.
(The green symbols would end up further to the left
if we assumed instead an age of $5\Gyr$.)
Here all stars are included---not just those for which
$P_{\rm rot}$ would also be available; see \Tab{TSum1}.
The inset shows $\tau/P_{\rm rot}^\ast$ as a function
of $B-V$ using \Eq{BV}.
The data points for the stars of M67 are overplotted to illustrate
the scatter and the range in $B-V$ covered by the data.
The red dotted line without surrounding data points shows the result
using the gyrochronology relation of \cite{Bar10} and \cite{BK10}
for $\tau(B-V)$, denoted by B+BK.

}\label{pRHK_4Gyr_all}\end{figure*}

Next, we compare with the diagram where $\tau/P_{\rm rot}^\ast$ is
estimated just from $B-V$ using gyrochronology; see \Eq{Pcyc_computed}
and \Fig{pRHK_4Gyr_all}.
Now, the direct fit for the 15 stars with
$\log\bra{R'_{\rm HK}}\ge-4.85$ gives
$\log c_2^{\rm dir}\approx-5.12$ and $\mu_2^{\rm dir}=-0.87$
and is shown as a dashed line.
The inset reveals that $\tau/P_{\rm rot}^\ast$ is indeed a monotonically
increasing function of $B-V$ in the range from $0.55$ to $0.8$,
as asserted earlier in this section.
The data points for the stars of M67 scatter around this line.
The corresponding relation obtained using the gyrochronology
relation of \cite{Bar10} is also given.
The difference of about $0.3$ dex results from the fact that the
$\tau(B-V)$ of \cite{BK10} is nearly twice as large as that of
\cite{Noyes84}.

As a function of $\tau/P_{\rm rot}^\ast$, the reversed trend of
$\log\bra{R'_{\rm HK}}$ is even more pronounced.
S1420 (green S) appears now more rapidly rotating:
$P_{\rm rot}^\ast=20.7\days$ whereas $P_{\rm rot}=24.8\days$;
see \Tab{TSum1}.
Another example is S1106 (green L) where
$P_{\rm rot}^\ast=24.3\days$ whereas $P_{\rm rot}=28.4\days$.
On the other hand, S801 (green C), S1218 (green N), and S1307
(green R) are now predicted to rotate slower than what is measured.
To understand these departures, we need to remind ourselves of the
possibility of measurement errors, notably in $P_{\rm rot}$, variability
of $\bra{R'_{\rm HK}}$ associated with cyclic changes in their
magnetic field, and of the intrinsically chaotic nature of
stellar activity.
Also, of course, the gyrochronology relation itself is only an
approximation to empirical findings and not a physical law of nature.

\section{Evolution and relation to reduced braking}
\label{Evolution}

Following \cite{vanSaders2016} and \cite{MvS17}, we would expect that
evolved stars lose their large-scale magnetic field and thereby
undergo reduced magnetic braking.
Their angular velocity should then stay approximately constant until
accelerated expansion occurs at the end of their main-sequence life.
For those stars, it might be difficult or even impossible to ever enter
the regime of antisolar DR.
This could be the case for $\alpha$~Cen~A (HD~128620, blue $k$),
KIC~8006161 (blue $o$), and 16 Cyg A and B (HD~186408 and 186427,
i.e., blue $q$ and $r$ symbols, respectively).
These are stars that rotate faster than expected based on their extremely
low chromospheric activity.
Given the intrinsic variability of stellar magnetic fields, it is
conceivable that the idea of reduced braking may not apply to all stars.
Others would brake sufficiently to enter the regime of antisolar
rotation and then exhibit enhanced activity, as discussed above.
With increasing age, those stars would continue to slow down further
and increase their chromospheric activity, as seen in \Fig{pRHK_4Gyr_all}.

\begin{figure*}[t!]\begin{center}
\includegraphics[width=\textwidth]{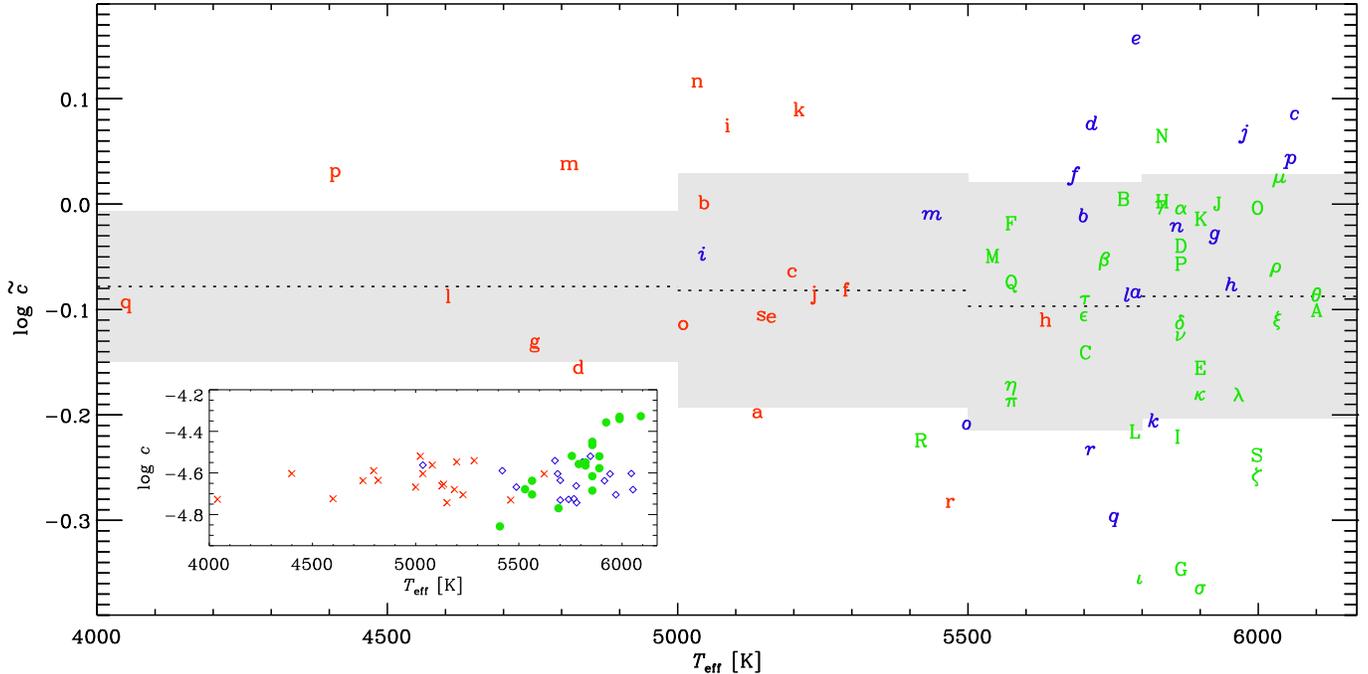}
\end{center}\caption[]{
Dependence of the residual $\log\tilde{c}$ on $\Teff$, which
corresponds to the dotted lines in \Figs{pRHK}{pRHK_4Gyr_all}.
Again, some of the symbols have been shifted to avoid overlapping.
Average and standard deviation are computed for smaller $\Teff$ intervals,
as indicated by horizontal dotted lines and gray boxes, respectively.
The inset shows the residual $\log c$ versus $\Teff$.
}\label{pRHK_comp}\end{figure*}

It is in principle possible that stars with different $\Teff$ show
a systematic dependence of the residual
\EQ
\log\tilde{c}=\log\bra{R'_{\rm HK}}
-\log\left[\;\mbox{``rhs of \Eq{representation}''}\;\right];
\label{cResidual3_tilde}
\EN
see the dotted lines in \Figs{pRHK}{pRHK_4Gyr_all}.
This is examined in \Fig{pRHK_comp}.
It turns out that this residual is essentially flat, i.e., there is
no systematic dependence on $\Teff$, and it is consistent with random
departures which do, however, becomes stronger toward larger $\Teff$,
as indicated by the gray boxes in \Fig{pRHK_comp}.

The work of \cite{KKKBOP15} has demonstrated that in the antisolar regime,
the magnetic activity can indeed be chaotic and intermittent.
Thus, depending on chance, a star in this regime may appear particularly
active (e.g., S1252, green O symbol with
$\log\bra{R'_{\rm HK}}=-4.72$), while others could be
particularly inactive (e.g., S969, green G symbol,
with $\log\bra{R'_{\rm HK}}=-5.06$).
Other examples are S1449 (green $\sigma$ with
$\log\bra{R'_{\rm HK}}=-5.13$) and S1048
(green $\iota$ with $\log\bra{R'_{\rm HK}}=-5.17$).
We must therefore expect that the magnetic activity of some of these
stars could still change significantly later in time, perhaps on
decadal or multi-decadal timescales.
In fact, we note from a comparison of the Ca~{\sc ii} measurements in
\cite{GBCSH17} with those from the initial chromospheric activity survey
of over a decade ago \citep{Giam2006} that the $R'_{\rm HK}$ values for
the specific stars mentioned above, S969 and S1048, are now each lower
by about 20\% while that for S1449 is lower by 23\%.

Given that the more massive stars of M67 are on their way to becoming
subgiants \citep[e.g.][]{Motta16}, we now discuss whether this could
explain their enhanced activity.
Properties important for convection such as luminosity and radius
may increase substantially above the main sequence values before reaching
the turnoff.
To compare with observations, it is convenient to look at the usual
residual $\log c=\log\bra{R'_{\rm HK}}-\log(\tau/P_{\rm rot})$, which
was given in the inset of \Fig{pRHK} as a function of $R'_{\rm HK}$ and
is now presented in the inset of \Fig{pRHK_comp} as a function of $\Teff$.
We see that the four hottest stars of the sample, S603 (green A),
S1095 (green J), S1252 (green O), and S1420 (green S) have a slight,
but systematic excess.
Assuming that their values of $R'_{\rm HK}$ and $P_{\rm rot}$
are accurate, this could mean that the estimated values of $\tau$
are too small.
\cite{Gil85} found that for a certain regime of evolution, stars of
the solar mass and above may have $\tau$ significantly larger (up to
$0.4\dex$) than those of main-sequence stars at the same effective
temperature (see their Figure~10).
However, the regime for this behavior occurred only when these stars
cooled to below the solar main-sequence effective temperature.
As can be seen in the color-magnitude diagram in \cite{Giam2006},
our sample does not include stars which have cooled to this degree; on the
contrary, our sample is still very near the main-sequence, and therefore
we expect \Eq{BV} should still apply.
This would therefore not alter our suggestion that most of the members
of M67 have antisolar DR.

\section{Conclusions}

The phenomenon of antisolar DR is well known from theoretical models of
solar/stellar convective dynamos in spherical shells.
So far, antisolar DR has only been observed in some K giants
\citep{SKW03,WSW05,Kovari_etal15,Kovari_etal17} and subgiants
\citep{Har16}, but not yet in dwarfs.
Our work is compatible with the interpretation that the 
enhanced activity at large Rossby numbers (slow rotation)
is a manifestation of antisolar DR.
Our results are suggestive of a bifurcation into two groups of stars:
those which undergo reduced braking and become inactive at
$P_{\rm rot}/\tau\approx2$ \citep{vanSaders2016}, and those that enter
the regime of antisolar rotation and continue to brake at enhanced
activity, although with chaotic time variability.
Interestingly, \cite{Katsova} have suggested that stars with antisolar
DR may be prone to exhibiting superflares \citep{Maehara,Candelaresi}.
This would indeed be consistent with the anticipated chaotic time
variability of such stars.

The available time series are too short to detect antisolar DR through
changes in the apparent rotation rate that would be associated with spots
at different latitudes; see \cite{RA15} for details of a new technique.
It is therefore important to use future opportunities, possibly still
with {\em Kepler}, to repeat those measurements at later times when the
magnetic activity belts might have changed in position.

\acknowledgements

We thank the referee for their thoughtful comments.
We are indebted to Bengt Gustafsson and Travis Metcalfe for useful
discussions and Dmitry Sokoloff for alerting us to their recent paper.
This work has been supported in part by
the NSF Astronomy and Astrophysics Grants Program (grant 1615100),
the Research Council of Norway under the FRINATEK (grant 231444),
and the University of Colorado through its support of the
George Ellery Hale visiting faculty appointment.
We gratefully acknowledge partial support of this investigation by grants
to AURA/NSO from, respectively, the NASA $Kepler/K2$ Guest Observer
program through Agreement No.\ NNX15AV53G and from the NN-EXPLORE program
through JPL RSA 1533727, which is administered by the NASA Exoplanet
Science Institute (NExScI).
The National Solar Observatory is operated by AURA under a cooperative
agreement with the National Science Foundation.

\newpage



\begin{thebibliography}{}

\bibitem[Baliunas et al.(1995)]{Bal+95}
Baliunas, S. L., Donahue, R. A., Soon, W. H., Horne, J. H., Frazer, J., Woodard-Eklund, L., Bradford, M., Rao, L. M., Wilson, O. C., Zhang, Q., Bennett, W., Briggs, J., Carroll, S. M., Duncan, D. K., Figueroa, D., Lanning, H. H., Misch, T., Mueller, J., Noyes, R. W., Poppe, D., Porter, A. C., Robinson, C. R., Russell, J., Shelton, J. C., Soyumer, T., Vaughan, A. H., \& Whitney, J. H.\yapj{1995}{438}{269}

\bibitem[Barnes(2010)]{Bar10}
Barnes, S. A.\yapj{2010}{722}{222}

\bibitem[Barnes \& Kim(2010)]{BK10}
Barnes, S. A., \& Kim, Y.-C.\yapj{2010}{721}{675}

\bibitem[Barnes et al.(2016)]{Barnes2016}
Barnes, S.~A., Weingrill, J., Fritzewski, D., Strassmeier, K.~G., \& Platais, I.\ 2016, \apj, 823, 16 

\bibitem[Brandenburg et al.(2017)]{BMM17}
Brandenburg, A., Mathur, S., \& Metcalfe, T. S.\yapj{2017}{845}{79}
(BMM)

\bibitem[Brown et al.(2011)]{BMBBT11}
Brown, B. P., Miesch, M. S., Browning, M. K., Brun, A. S., \& Toomre, J.\yapj{2011}{731}{69}

\bibitem[Candelaresi et al.(2014)]{Candelaresi}
Candelaresi, S., Hillier, A., Maehara, H., Brandenburg, A., \& Shibata, K.\yapj{2014}{792}{67}

\bibitem[Fan \& Fang(2014)]{FF14}
Fan, Y., \& Fang, F.\yapj{2014}{789}{35}

\bibitem[Gastine et al.(2014)]{GYMRW14}
Gastine, T., Yadav, R. K., Morin, J., Reiners, A., \& Wicht, J.\ymn{2014}{438}{L76}

\bibitem[Giampapa et al.(2017)]{GBCSH17}
Giampapa, M. S., Brandenburg, A., Cody, A. M., Skiff, B. A., \& Hall, J. C.\sapjE{2017}
{http://www.nordita.org/preprints, no.~2017-121}

\bibitem[Giampapa et al.(2006)]{Giam2006}
Giampapa, M. S., Hall, J. C., Radick, R. R., \& Baliunas, S. L.\yapj{2006}{651}{444}

\bibitem[Gilliland(1985)]{Gil85}
Gilliland, R. L.\yapj{1985}{299}{286}

\bibitem[Gilman(1977)]{Gil77}
Gilman, P. A.\ygafd{1977}{8}{93}

\bibitem[Harutyunyan et al.(2016)]{Har16}
Harutyunyan, G., Strassmeier, K. G., K\"unstler, A., Carroll, T. A., \& Weber, M.\yana{2016}{592}{A117}

\bibitem[K\"apyl\"a et al.(2014)]{KKB14}
K\"apyl\"a, P. J., K\"apyl\"a, M. J., \& Brandenburg, A.\yana{2014}{570}{A43}

\bibitem[Karak et al.(2015)]{KKKBOP15}
Karak, B. B., K\"apyl\"a, M. J., K\"apyl\"a, P. J., Brandenburg, A., Olspert, N., \& Pelt, J.\yana{2015}{576}{A26}

\bibitem[Katsova et al.(2018)]{Katsova}
Katsova, M. M., Kitchatinov, L. L., Livshits, M. A., Moss, D. L., Sokoloff, D. D., \& Usoskin, I. G.\yarep{2018}{95}{78}

\bibitem[K{\H o}v{\'a}ri et al.(2015)]{Kovari_etal15}
K{\H o}v{\'a}ri, Z., Kriskovics, L., K\"unstler, A., Carroll, T. A., Strassmeier, K. G., Vida, K., Ol\'ah, K., Bartus, J., Weber, M.\yana{2015}{573}{A98}

\bibitem[K{\H o}v{\'a}ri et al.(2017)]{Kovari_etal17}
K{\H o}v{\'a}ri, Z., Strassmeier, K. G., Carroll, T. A., Ol\'ah, K., Kriskovics, L., K{\H o}v{\'a}ri, E., Kov\'acs, O., Vida, K., Granzer, T., \& Weber, M.\yana{2017}{606}{A42}

\bibitem[Maehara et al.(2012)]{Maehara}
Maehara, H., Shibayama, T., Notsu, S., Notsu, Y., Nagao, T., Kusaba, S., Honda, S., Nogami, D., \& Shibata, K.\ynat{2012}{485}{478}

\bibitem[Mamajek \& Hillenbrand(2008)]{MH08}
Mamajek, E. E., \& Hillenbrand, L. A.\yapj{2008}{687}{1264}

\bibitem[Metcalfe \& van Saders(2017)]{MvS17}
Metcalfe, T. S., \& van Saders, J.\ysph{2017}{292}{126}

\bibitem[Motta et al.(2016)]{Motta16}
Motta, C. B., Salaris, M., Pasquali, A., \& Grebel, E. K.\ymn{2016}{466}{2161}

\bibitem[Noyes et al.(1984)]{Noyes84}
Noyes, R. W., Hartmann, L., Baliunas, S. L., Duncan, D. K., \& Vaughan, A. H.\yapj{1984}{279}{763}

\bibitem[{\"O}nehag et al.(2011)]{Onehag2011}
{\"O}nehag, A., Korn, A., Gustafsson, B., Stempels, E., \& Vandenberg, D.~A.\ 2011, \aap, 528, A85

\bibitem[Reinhold \& Arlt(2015)]{RA15}
Reinhold, T., \& Arlt, R.\yana{2015}{576}{A15}

\bibitem[Saar \& Brandenburg(1999)]{SB99}
Saar, S. H., \& Brandenburg, A.\yapj{1999}{524}{295}
(SB)

\bibitem[Sarajedini et al.(2009)]{Sara2009}
Sarajedini, A., Dotter, A., \& Kirkpatrick, A.\ 2009, \apj, 698, 1872

\bibitem[Strassmeier et al.(2003)]{SKW03}
Strassmeier, K. G., Kratzwald, L., \& Weber, M.\yana{2003}{408}{1103}

\bibitem[van Saders et al.(2016)]{vanSaders2016}
van Saders, J. L., Ceillier, T., Metcalfe, T. S., Silva Aguirre, V., Pinsonneault, M. H., Garc\'ia, R. A., Mathur, S., \& Davies, G. R.\ynat{2016}{529}{181}

\bibitem[Vilhu(1984)]{Vil84}
Vilhu, O.\yana{1984}{133}{117}

\bibitem[Weber et al.(2005)]{WSW05}
Weber, M., Strassmeier, K. G., \& Washuettl, A.\yan{2005}{326}{287}

\end{thebibliography}
\end{document}